\documentclass{acmsiggraph}
\DeclareOldFontCommand{\bf}{\normalfont\bfseries}{\mathbf}

\usepackage[table]{xcolor}
\usepackage{tabularx}
\usepackage{booktabs}
\usepackage{pdfcomment}
\usepackage{amsmath}

\usepackage{amsfonts}
\usepackage{amssymb}

\usepackage[english]{babel}

\definecolor{lightbluishgrey}{rgb}{0.78,0.86,0.93}

\definecolor{highlightgreen}{rgb}{0.1,0.7,0.1}

\newcommand{\reviewer}[1]{{}} 

\newcommand{\LK}[1]{}

\newcommand{\cheatvspace}[1]{}

\newcommand{\PWN}{}
\def\PWN/{PWN}
\newcommand{\numpwn}{}
\def\numpwn/{8616}

\usepackage{layouts}

\usepackage{wrapfig}

\newcommand{\reffig}[1] {Figure~\ref{fig:#1}}

\usepackage{xfrac}

\usepackage{amsmath}
\usepackage{amssymb}    
\usepackage{cancel}
\usepackage[T1]{fontenc}

\usepackage{siunitx}
\usepackage[mathletters]{ucs}
\usepackage[utf8x]{inputenc}
\usepackage{upgreek}
\usepackage{dblfloatfix}

\usepackage{graphicx}

\usepackage{wrapfig}

\makeatletter
\let\save@mathaccent\mathaccent
\newcommand*\if@single[3]{%
  \setbox0\hbox{${\mathaccent"0362{#1}}^H$}%
  \setbox2\hbox{${\mathaccent"0362{\kern0pt#1}}^H$}%
  \ifdim\ht0=\ht2 #3\else #2\fi
  }
\newcommand*\rel@kern[1]{\kern#1\dimexpr\macc@kerna}
\newcommand*\widebar[1]{\@ifnextchar^{{\wide@bar{#1}{0}}}{\wide@bar{#1}{1}}}
\newcommand*\wide@bar[2]{\if@single{#1}{\wide@bar@{#1}{#2}{1}}{\wide@bar@{#1}{#2}{2}}}
\newcommand*\wide@bar@[3]{%
  \begingroup
  \def\mathaccent##1##2{%
    \let\mathaccent\save@mathaccent
    \if#32 \let\macc@nucleus\first@char \fi
    \setbox\z@\hbox{$\macc@style{\macc@nucleus}_{}$}%
    \setbox\tw@\hbox{$\macc@style{\macc@nucleus}{}_{}$}%
    \dimen@\wd\tw@
    \advance\dimen@-\wd\z@
    \divide\dimen@ 3
    \@tempdima\wd\tw@
    \advance\@tempdima-\scriptspace
    \divide\@tempdima 10
    \advance\dimen@-\@tempdima
    \ifdim\dimen@>\z@ \dimen@0pt\fi
    \rel@kern{0.6}\kern-\dimen@
    \if#31
      \overline{\rel@kern{-0.6}\kern\dimen@\macc@nucleus\rel@kern{0.4}\kern\dimen@}%
      \advance\dimen@0.4\dimexpr\macc@kerna
      \let\final@kern#2%
      \ifdim\dimen@<\z@ \let\final@kern1\fi
      \if\final@kern1 \kern-\dimen@\fi
    \else
      \overline{\rel@kern{-0.6}\kern\dimen@#1}%
    \fi
  }%
  \macc@depth\@ne
  \let\math@bgroup\@empty \let\math@egroup\macc@set@skewchar
  \mathsurround\z@ \frozen@everymath{\mathgroup\macc@group\relax}%
  \macc@set@skewchar\relax
  \let\mathaccentV\macc@nested@a
  \if#31
    \macc@nested@a\relax111{#1}%
  \else
    \def\gobble@till@marker##1\endmarker{}%
    \futurelet\first@char\gobble@till@marker#1\endmarker
    \ifcat\noexpand\first@char A\else
      \def\first@char{}%
    \fi
    \macc@nested@a\relax111{\first@char}%
  \fi
  \endgroup
}
\makeatother

\usepackage[ruled,vlined]{algorithm2e}
\usepackage{etoolbox}

  \makeatletter
  \patchcmd{\algocf@Vline}{\vrule}{\vrule\hspace{-0.25em}}{}{}
  \makeatother
  \DontPrintSemicolon
  \SetAlgoLined
  \SetKwInput{KwData}{Inputs} 
  \SetKwInput{KwResult}{Outputs}
  \RestyleAlgo{ruled}
  \SetKwBlock{Repeat}{repeat}{}
 
\usepackage{array}

\makeatletter
\newcommand{\raisemath}[1]{\mathpalette{\raisem@th{#1}}}
\newcommand{\raisem@th}[3]{\raisebox{#1}{$#2#3$}}
\makeatother

\newcommand{\thetitle}{}
\def\thetitle/{Mesh Arrangements for Solid Geometry}

\usepackage{enumitem}
\usepackage{hyperref}
\hypersetup{
    pdfinfo={
        Title={Thingi10K: A Dataset of 10,000 3D-Printing Models},
        Author={Qingnan Zhou, Alec Jacobson},
        Subject={},
        Keywords={}
    }
}

\usepackage{t1enc,dfadobe}

\title{
Thingi10K: A Dataset of 10,000 3D-Printing Models
}

\author{
Qingnan Zhou\\
New York University
\and
Alec Jacobson\\
Columbia University
}
\pdfauthor{Qingnan Zhou, Alec Jacobson}

\begin{document}

\teaser{
\includegraphics[width=\linewidth]{figs/thingi-dataset.pdf}
\caption{The Thingi10K dataset contains 10,000 models from from featured
``things'' on thingiverse.com, a popular online repository.}
\label{fig:10k}
}

\maketitle

\begin{abstract}
Empirically validating new 3D-printing related algorithms and implementations requires testing data representative of inputs encountered \emph{in the wild}. An ideal benchmarking dataset should not only draw from the same distribution of shapes people print in terms of class (e.g., toys, mechanisms, jewelry), representation type (e.g., triangle soup meshes) and complexity  (e.g., number of facets), but should also capture problems and artifacts endemic to 3D printing models (e.g., self-intersections, non-manifoldness). We observe that the contextual and geometric characteristics of 3D printing models differ significantly from those used for computer graphics applications, not to mention standard models (e.g., Stanford bunny, Armadillo, Fertility). We present a new dataset of 10,000 models collected from an online 3D printing model-sharing database. Via analysis of both geometric (e.g., triangle aspect ratios, manifoldness) and contextual  (e.g., licenses, tags, classes) characteristics, we demonstrate that this dataset represents a more concise summary of real-world models used for 3D printing compared to existing datasets. To facilitate future research endeavors, we also present an online query interface to select subsets of the dataset according to project-specific characteristics. The complete dataset and per-model statistical data are freely available to the public.

\end{abstract}

\section{Introduction and background}
\label{sec:intro}

The iconic \emph{Stanford bunny}, now 23 years old, has been melted, shattered,
and deformed countless times.  While mostly a fun subculture, ``bunny torture''
is also a legacy of an earlier time when few interesting and free 3D models
existed.
Testing on such standard models persists despite well-known limitations.
As Greg Turk, originator of the bunny, advises, ``I actually consider the bunny
to be \emph{too good} as a test model. It is fairly smooth, it has manifold
connectivity, and it isn't too complex'' \cite{stanfordbunny}.

Oversimplified testing provides a false sense of robustness and
causes not only visual artifacts in computer
graphics applications, but also fabrication and functionality artifacts when
processing geometry intended for 3D printing.
Fortunately, 3D models are now abundant.
Modern consumer-level 3D printing technologies nurture new communities of
professional and amateur 3D modelers, who share and sell 3D-printable models
online (e.g., \url{shapeways.com}, \url{sketchfab.com}, \url{thingiverse.com}).
This wealth of data also echoes the demand for state-of-the-art
processing techniques and automation within 3D printing pipelines.

However, testing remains inadequate. 
Existing datasets contain only sanitized models (e.g.,
\cite{aimatshape,levoy2005,MPZ14}) or draw from populations containing raw
models not specifically intended for printing (rather, e.g., for shape
classification \cite{shilane2004,shapenet2015} or scene understanding
\cite{Silberman:ECCV12,Choi2016}).

In this paper, we will show that the characteristics and issues common to 3D
printing models are distinct from models intended for visualization.
As such, validating geometry processing techniques related to 3D printing
requires a new representative dataset.
This ideal dataset should encompass the different contextual and geometric
characteristics of commonly printed shapes.
Characteristics common to 3D printing models should appear with proportional
distributions, and characteristics \emph{inconsistent} with models intended for
fabrication should be infrequent (e.g., the open boundaries of a video game
character's clothing).

We propose a dataset of 10,000 models culled from a popular shape repository
for 3D printing enthusiasts, \url{thingiverse.com}.
Hereon, we refer to our dataset as \emph{Thingi10K}.
Beyond collecting tags and class information available online, we analyze
geometric characteristics of each model (e.g., manifoldness, lack of
self-intersections, genus).
We contrast these statistics against existing large datasets and investigate
correlations within the data.

\newcommand{\parag}[1]{\textbf{#1}$\quad$}

\parag{Existing datasets.}
Myles et al.\ collect 116 models from academic sources (Stanford Scanning
Repository \cite{levoy2005} and Aim@Shape Repository \cite{aimatshape}) to test
their parameterization algorithm \cite{MPZ14}.
These models correspond to \emph{best-case} input due to their extreme
cleanliness and general position assumption (i.e., no four points on a
circle, no coplanar intersections, etc.).
For 3D printing models in the wild,  degeneracies, non-manifoldness and
self-intersections are abundant, not special cases. 
Structured modeling and coordinate quantization tends to break rather than
fulfill general position assumptions.

\begin{figure}[th]
\centering
\includegraphics[width=1.0\columnwidth]{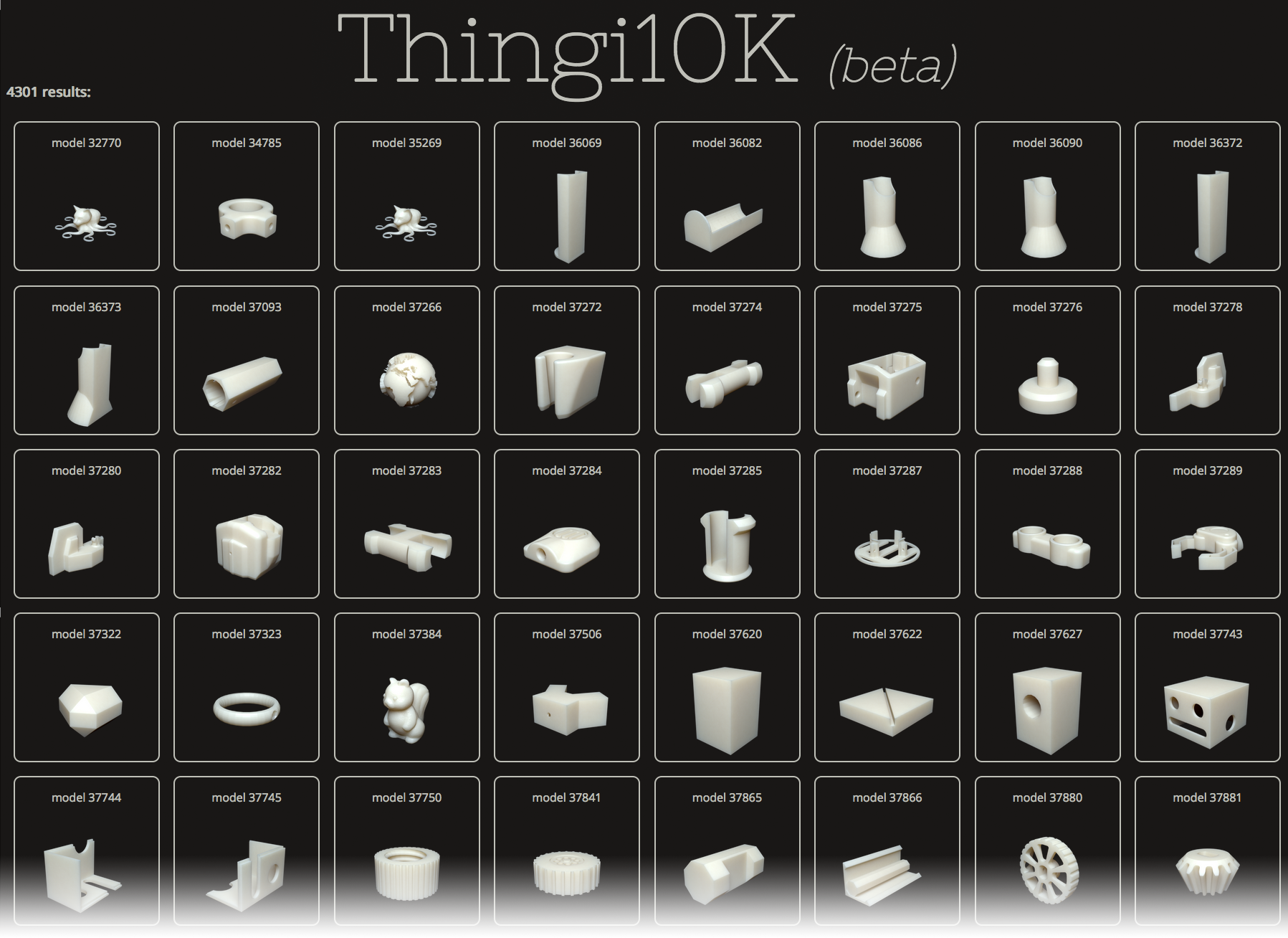}
\caption{Our online query interface selects subsets of Thingi10K.}
\label{fig:online_ui_result}
\end{figure}

Computer vision and machine learning applications demand large scale training
datasets.
For example, the NYU Depth Dataset collects thousands of depth video sequences
of indoor scenes for object classification \cite{Silberman:ECCV12}.
The Princeton Shape Benchmark collects 1,814 polygonal models of specific
objects (e.g., animals, furniture) from various internet sources for shape
classification \cite{shilane2004}.
More recently, ShapeNet collects more than three million annotated models
\cite{shapenet2015}.
The ShapeNetCore subset contains 57,459 single-object models with
semi-automatically generated category information.
Although models from these datasets resemble physical objects, their geometric
characteristics suggest their intention was for visualization rather than
fabrication.
These datasets are not suitable for testing 3D printing techniques.

In addition to generic datasets, a variety of specialized datasets exist.
For example, Lim et al.\ provide 219 IKEA 3D models for
pose-estimation \cite{lpt2013ikea}.
Recently, Choi et al.\ released a dataset of 10,000 scanned objects, with a subset
of 383 successfully reconstructed 3D models \cite{Choi2016}.
The Shape Retrieval Contest releases multiple datasets each year
to test retrieval algorithms including generic
\cite{bronstein2010shreclarge,li2012shrec}, non-rigid humans
\cite{pickup2014shrec}, sketch-based shapes \cite{li2013shrec,li2014shrec},
shape correspondences \cite{bronstein2010shrec}, facial expressions
\cite{nair2008shrec,veltkamp2011shrec}, and range scans
\cite{dutagaci2010shrec}.
Our Thingi10K dataset complements these sources by providing a specialized
dataset for 3D printing objects.

We are not the first to utilize Thingiverse models for academic purposes. 
To test a rapid prototyping interface, Mueller et al.\ consider Thingiverse
models, but report that meshing artifacts required manual cleanup before
processing \cite{Mueller:2014:FFP}. 
Beyer et al.\ procedurally collect 2,250 models with specific tags from
Thingiverse to test a decomposition algorithm \cite{Beyer:2015}.
Buehler et al.\ manually sift through 25,000 models from search results
on Thingiverse to identify 363 models as ``assistive technologies''
\cite{Buehler:2015:SCA}.
Beyond testing a specific routine, these works do not analyze low-level
geometric characteristics of the collected models.
These \emph{collected} datasets are also not publicly available.

\parag{Contributions.}
Unlike previous datasets, our Thingi10K dataset reflects
the variety, complexity and (lack of) quality of 3D printing models.
It is immediately useful for testing the performance of methods for
structural analysis \cite{Stava:2012,Zhou2013,Umetani:2013}, shape optimization
\cite{Prevost:2013,Bacher:2014,Musialski:2015}, or solid geometry operations
\cite{zhou2016}.
Due to its specialized nature and correlated contextual information, we suspect
the dataset is also useful for machine learning and data mining algorithms.
We compare the collected contextual information and computed geometric
properties of our dataset in detail against two existing datasets: MPZ14 and
ShapeNetCore.
We demonstrate that these represent two extreme cases
in terms geometric quality while our dataset provides a mixture of geometric
qualities reflecting real-world settings.
All data and analysis of our dataset are freely available to the public.
To facilitate exploration and future reuse, we provide an
easy-to-use online query interface (see \reffig{online_ui_result}).
This interface augments the Thingiverse front-end with our geometric
analysis of each model.

\begin{figure}[th]
\centering
\includegraphics[width=\columnwidth]{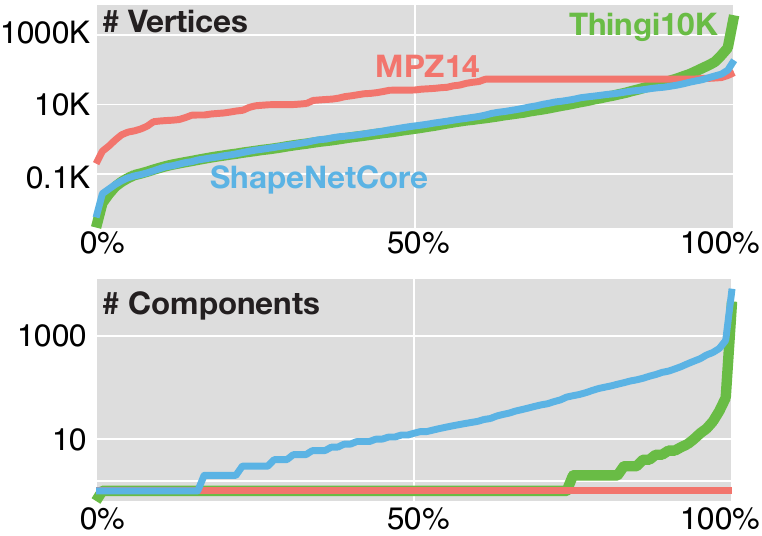}
\caption{Percentile plots of vertex and component count.}
\label{fig:complexity}
\end{figure}

\section{Methodology}
\label{sec:method}
Instead of our hiring professional modelers or scanning physical objects, we
leverage the availability of 3D models hosted and shared online.  Among all 3D
shape repositories, we select Thingiverse for its large and active user
community, its vast collection of print-validated designs, and its restriction
to open-source licenses.

As one of the largest online shape repositories, Thingiverse hosts more than a
million user-uploaded \emph{things},
3D designs consisting of one or more 3D \emph{models} (i.e., one or more mesh
files).
As of October 2015, Thingiverse has more than 2 million active 
users,
with 30-40 uploads each week and 1.7 million downloads per month \cite{makerbot1M}.
Thanks to this community, a design is typically not only modeled virtually but
also fabricated by one or more users, which provides invaluable real-world
validations.

Our Thingi10K dataset consists of 10,000 models (from 2011 things)
systematically culled from Thingiverse via web crawling.
Rather than randomly sample the entire repository, which may contain
bogus models uploaded by inexperienced users or for testing purposes,
we focus on things \emph{featured} on Thingiverse.
Featured things are entirely and independently selected by Thingiverse staff
based on their design, beauty and manufacturability. In a sense, these 10,000
models represent a subset of the top-quality designs on Thingiverse.
Thingi10K contains every 3D model of every thing featured by Thingiverse
between Sept. 16, 2009 and Nov. 15, 2015.

\section{Analysis}
\label{sec:analysis}

The 10,000-model dataset comes from 2,011 unique things designed by 1,083 unique
users, covering a large variety. Nearly all models are stored as .stl files
(9,956); the rest are .obj (42), .ply (1), and .off (1).
We analyze both geometric and contextual information of our dataset to
illustrate its representational quality and diversity.

\subsection{Geometry information}
\label{subsec:geo_analysis}

We analyze a variety of mesh complexity and quality measures on our dataset of
3D printing models and compare with two existing datasets: MPZ14 (116 models)
and ShapeNetCore (2000 models uniformly sampled from 57,459).

\subsubsection{Complexity} 
Complexity of 3D model does not directly correlate with 3D printing cost.
We evaluated three different measures to quantify the complexity of our
dataset: number of vertices, number of disconnected components and genus.

\begin{figure}
\centering
\includegraphics[width=\columnwidth]{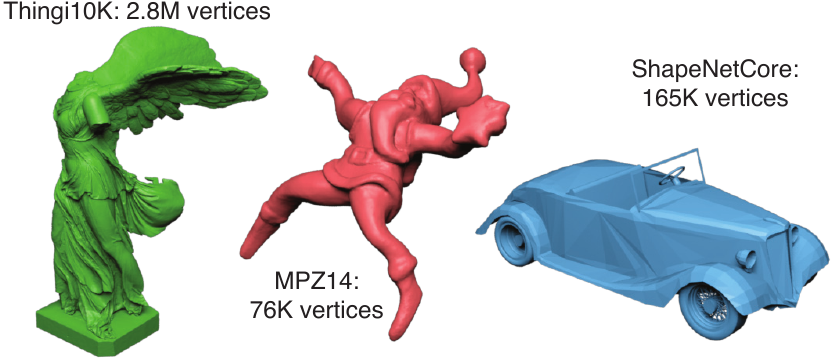}
\caption{Highest resolution models from each dataset.}
\label{fig:high_res_models}
\end{figure}

Figure \ref{fig:complexity} provides the percentile plot of both vertex and
component count over each dataset.  The vertex count plot indicates that the
MPZ14 dataset favors moderately high resolution models and excludes extremely
low or high resolution models.  On the other hand, the distribution over our
dataset and ShapeNetCore is similar, with our dataset covering a larger range.
Figure \ref{fig:high_res_models} illustrates the highest resolution model of
each dataset.

\begin{figure}
\centering
\includegraphics[width=\columnwidth]{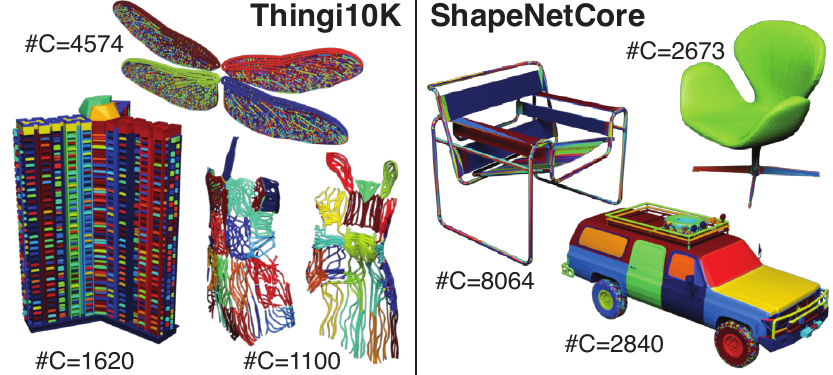}
\caption{Connected components of Thingi10K models tend to represent salient
parts; those in ShapeNetCore are often just disconnected patches
(models with most components shown).}
\label{fig:multi_comp_meshes}
\end{figure}

Many geometry processing algorithms assume input will be processed one
component at a time, so it is not a surprise that MPZ14 contains exclusively
single-component models.
This assumption is not valid in the context of 3D printing,
where multiple components could overlap to form a larger shape.
Analysing each component separately may lead to incorrect results.
Within our dataset 29\% of models have more than one component. ShapeNetCore
is 83\% multi-component, but close inspection finds many models are composed of
incoherent patches or isolated faces (see \reffig{multi_comp_meshes})
In contrast, 3D printing models with high numbers of components in Thingi10K
are typically by design, with the base shape naturally decomposing into smaller
components.

\begin{figure}
\centering
\includegraphics[width=\columnwidth]{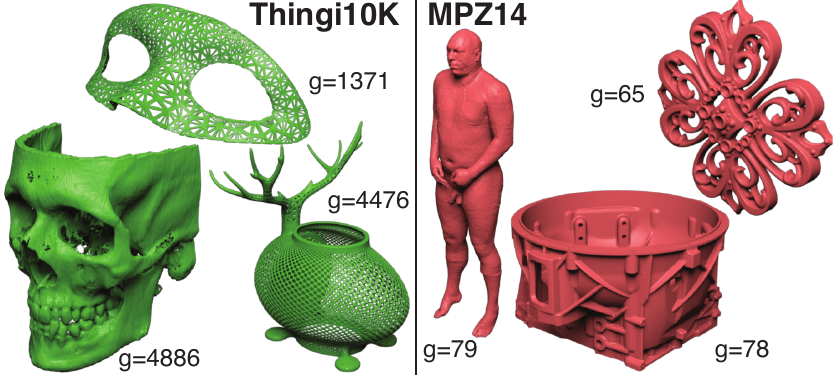}
\caption{Models with the highest genus from each dataset.}
\label{fig:high_genus_examples}
\end{figure}

The genus distribution of our Thingi10K dataset is similar to MPZ14, but our
dataset covers a larger range of genus, with the highest genus over 60 times
larger than in MPZ14 (see Figure \ref{fig:high_genus_examples}).
To avoid confusion, we limit the genus comparison to single-component, closed
and manifold meshes. Zero of the ShapeNetCore models meet this criteria.

\subsubsection{Mesh quality}
Mesh qualities of a dataset play a major role in determining its usability and
representation of models \emph{in the wild}.
For example, degenerate or sliver triangles will cause poor accuracy in
non-robust finite element simulations, and fragile volumetric meshing routines
will fail in the presence of self-intersections.
It is crucial to understand the mesh quality of real-world input data in order
to design robust and practical algorithms.
Existing datasets often focus on high-level properties and provide little
insight on their mesh qualities.  Our analysis aims to fill this gap. 

\begin{figure}
\centering
\includegraphics[width=\columnwidth]{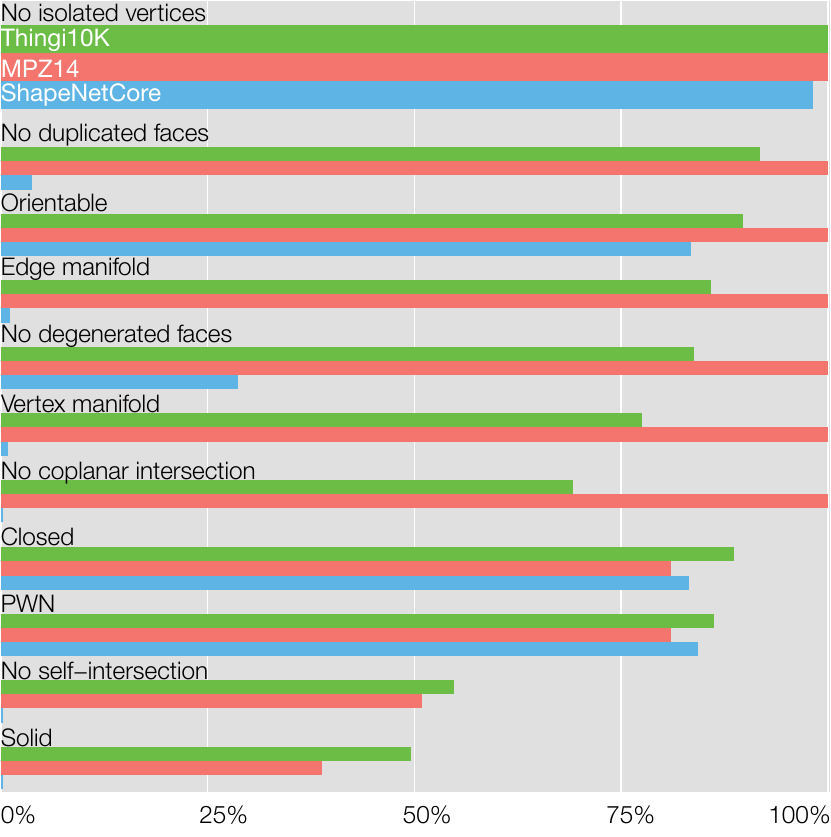}
\caption{MPZ14 models are ``too clean,'' whereas ShapeNetCore are
unrealistically corrupted in the context of 3D printing.}
\label{fig:input_quality_comp}
\end{figure}

We analyze 13 mesh quality measurements:

\begin{enumerate}[style=unboxed,labelwidth=0cm,labelsep=0cm,leftmargin=0cm]
\item[] \textbf{Closed}: Every edge is adjacent to 2 or more faces.

\item[] \textbf{Oriented}: Every non-boundary edge has
zero signed incidence. In other words, the number of positively oriented incident
faces must equal to the number of negatively oriented incident faces.

\item[] \textbf{No isolated vertices}: All vertices are adjacent to at least one face.

\item[] \textbf{No duplicated faces}: There does not
exist a pair of faces sharing the same set of vertices.

\item[] \textbf{Vertex-manifold}: The one-ring neighborhood of
every vertex is a topological disc.

\item[] \textbf{Edge-manifold}: Every non-boundary edge must be incident to exactly two
faces.

\item[] \textbf{No degeneracy}: All faces must have
non-collinear vertices.
Degeneracy can be checked with exact predicates \cite{shewchuk1997}.

\item[] \textbf{No self-intersection}: The intersection of any two faces is either empty, a shared vertex, or a shared edge.
Exact predicates are necessary to ensure correctness.

\item[] \textbf{No coplanar intersections}: No two faces are coplanar and overlapping.  
This is a strictly weaker condition than ``no self-intersection.''

\item[] \textbf{Piecewise-constant winding number (PWN)}: The winding
number field at any non-mesh point is piece-wise constant (\cite{zhou2016}).

\item[] \textbf{Solid}: The input mesh must be a valid boundary of a
subspace of $\mathbb{R}^3$.  Specifically, it must be PWN, self-intersection free and
induce a \{0, 1\} winding number field.

\item[] \textbf{Aspect ratio}: The aspect ratio of a triangle is the ratio of
its circumradius to the diameter of its incircle.

\item[] \textbf{Intrinsically Delaunay}: All edges must have non-negative
cotangent weights \cite{fisher2007}.
\end{enumerate}

These mesh quality measures are not by no means complete.
Additional quality measures (\cite{shewchuk2002,attene2013})
can be easily adopted.

\begin{figure}[th]
\includegraphics[width=\columnwidth]{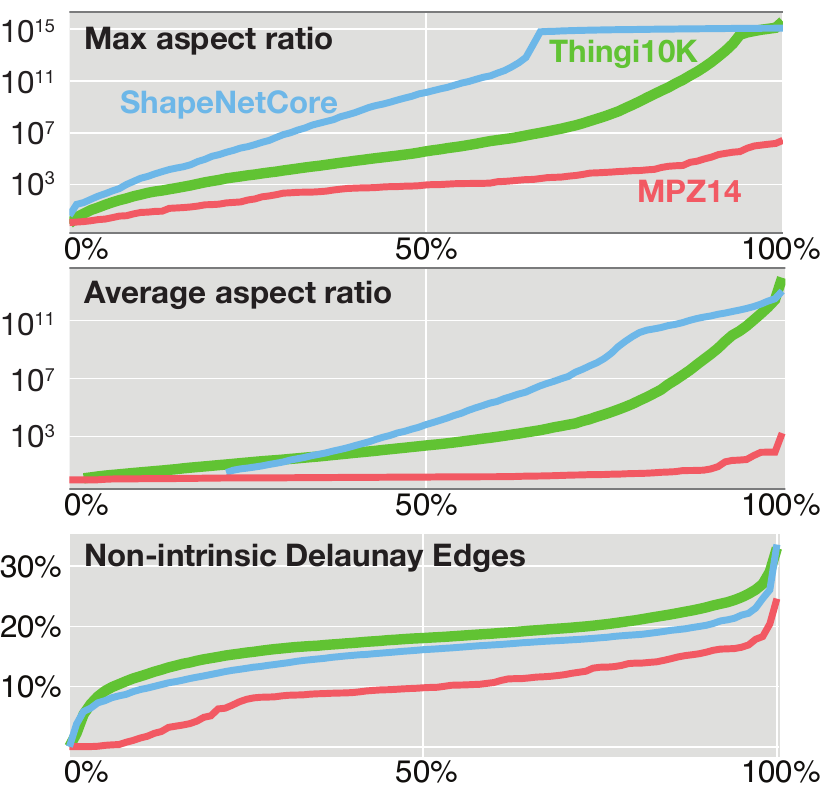}
\caption{Percentile plots mesh quality measures.}
\label{fig:aspect_ratio}
\end{figure}

\begin{figure*}[h]
\includegraphics[width=\textwidth]{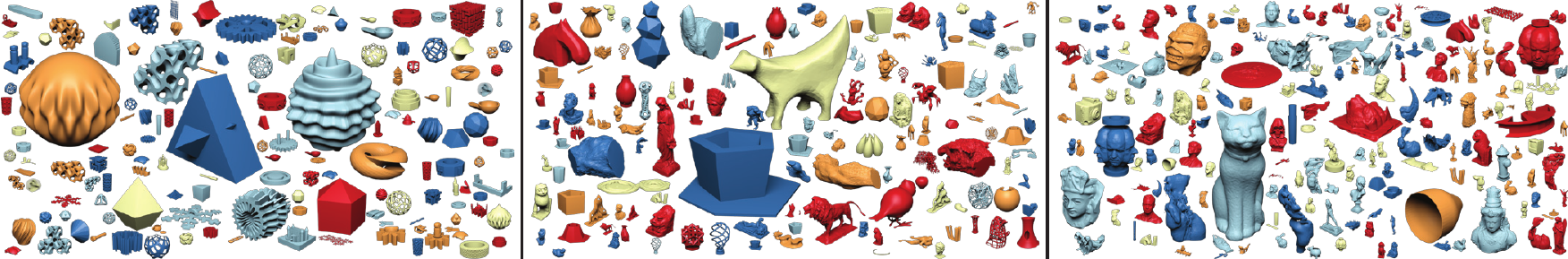}
\caption{Models with tag \texttt{math} (left), \texttt{sculpture} (middle) and
\texttt{scan} (right). }
\label{fig:tag_examples}
\end{figure*}

Figure \ref{fig:input_quality_comp} shows the percentage of models that
satisfy each of the first 11 quality measures.  Figure \ref{fig:aspect_ratio}
illustrates the maximum, average aspect ratio and
the fraction of non-intrinsic Delaunay edges over all models in each dataset.
Our analysis shows that MPZ14 has ``unrealistically pristine'' mesh quality,
whereas ShapeNetCore exhibits mesh quality issues not common to 3D printed
models, reflecting that it is gathered from a larger space of 3D models.

MPZ14 has \emph{perfect} mesh quality according to seven different measures.  In
particular, all models are manifold, oriented and degeneracy-free.  Because
many geometry processing algorithms do not require the input model to be closed
or self-intersection free, data from MPZ14 are perfect as proof-of-concept examples.
However, their high quality is due to the fact that models were selected not on
merits of their shape, functionality or aesthetics, but rather because they meet
certain quality criteria or have been sanitized.

\begin{figure}[h]
\centering
\includegraphics[width=1.0\columnwidth]{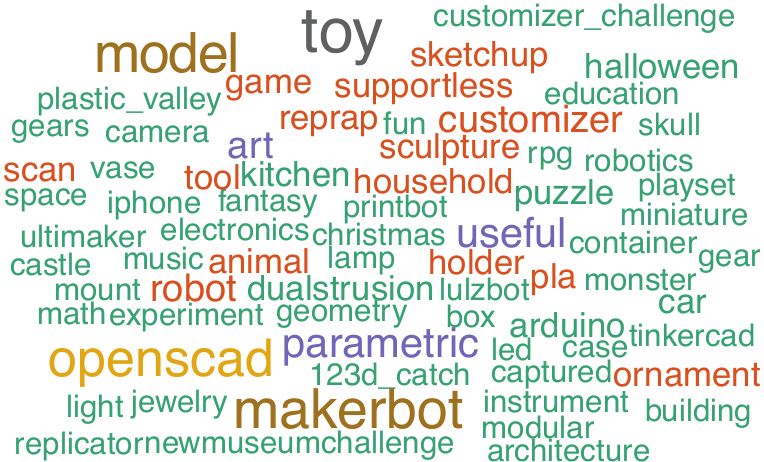}
\caption{Thingi10K user tags highlight the dataset's variety.}
\label{fig:tag_cloud}
\end{figure}

On the other hand, ShapeNetCore has very poor mesh quality according to 6
measures in Figure \ref{fig:input_quality_comp}.
Its maximum and average triangle aspect ratios are visibly worse than Thingi10K
and MPZ14.
This is partially due to the fact that these data are collected directly
from the internet, where models were not necessary designed for fabrication
purposes.
Many existing learning algorithms side-step the quality issues by transforming
boundary representations to depth images or bounding box hierarchies
\cite{hu2012}.
Performing geometry processing algorithms directly on these models is very hard
due to poor mesh quality.

In contrast, our dataset offers a curated collection of 3D meshes with a large
range of mesh qualities.  It contains a significant number of high quality
models as well as a non-negligible proportion of models with common mesh quality
problems. Due to its large quantity, our dataset is ideal for stress-testing
purposes where one can easily select a subset of the data that matches any
combination of mesh criteria (Section \ref{sec:interface}). Because all data
are sampled from real-world models designed to be 3D printed, our dataset
provides an unbiased view of the mesh qualities used in practice.  Our analysis
could be used to gauge the restrictions posed by various assumptions on mesh
quality. For example, an algorithm assuming self-intersection-free input would
automatically exclude 45\% of inputs, which may not be acceptable in
a real-world settings.

\subsection{Contextual information}

Each thing in our dataset is annotated by its original
designer.  Thingiverse supports three types of annotations: category,
subcategory and tags.  The first two must be selected from a predefined list of
categories, and the last one is a set of free-form texts created by the user.
A total of 4892 distinct tags are used in our dataset.  Figure
\ref{fig:tag_cloud} illustrates the most frequently used tags.

Unlike ShapeNet \cite{shapenet2015}, which focuses on providing
categorical annotations specific to object classification purposes, our
dataset comes with a rich and diverse set of original tags ranging from the
semantics of a 3D model to the printer/material used for fabrication.  For
example, Figure \ref{fig:tag_examples} shows all models with tags
\texttt{math}, \texttt{sculpture} and \texttt{scan}.

When combined with geometric analysis, our annotations reveal interesting
insights unavailable from previous works. For example, a simple frequency
analysis indicates OpenSCAD is the most popular modeling tool used by
Thingiverse users.  Our dataset shows that 98\% of OpenSCAD models are closed,
while only 91\% of SketchUp models and 85\% of TinkerCAD models are closed

\begin{figure}[h]
\centering
\includegraphics[width=1.0\columnwidth]{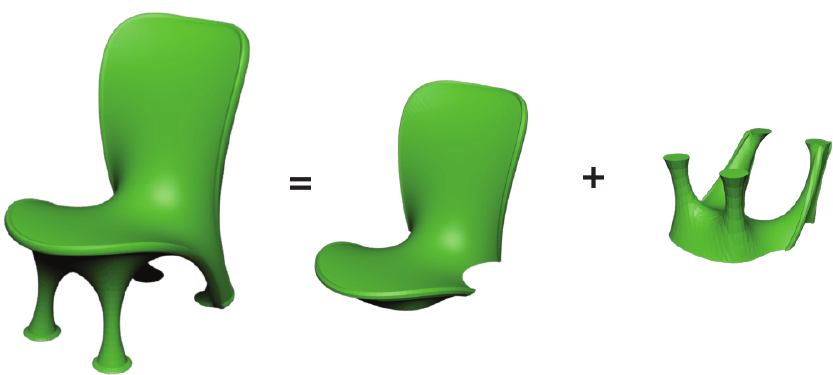}
\caption{A soap bubble chair is decomposed and re-oriented by its designer for
support-free 3D printing.}
\label{fig:decompose}
\end{figure}

Furthermore, due its fabrication-focused nature, many uploaded meshes are
``print-ready'' in the sense that their orientation and decompositions are designed for
optimal printing outcome (See figure \ref{fig:decompose}).
Recent papers have tried to solve
problems such as decomposing a large model to fit in the print volume and
finding ideal
print orientations \cite{Chen:2015}. The Thingi10K models, by their intrinsic nature of being
successful prints, represent ground truth data.

\begin{figure}[bh]
\centering
\includegraphics[width=1.0\columnwidth]{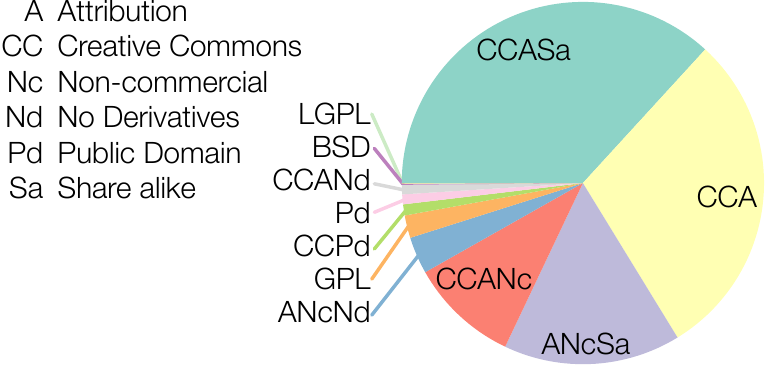}
\caption{All 10,000 models come under open source licenses.}
\label{fig:license}
\end{figure}

Lastly, all things are published under one of the open source licenses.
Figure \ref{fig:license} illustrates all licenses supported by Thingiverse.

\section{Online query interface}
\label{sec:interface}

To facilitate our goal of understanding 3D printed shapes, we provide an
online query interface, \url{ten-thousand-models.appspot.com} for
anyone to explore and dissect the dataset. The query terms may consist of one or
more clauses.  Each clause specifies a single search condition, e.g.
``genus>100''. Multiple clauses are separated by commas, and the search engine
retrieves models that satisfy all search conditions.

\begin{figure}[tbh]
\centering
\includegraphics[width=1.0\columnwidth]{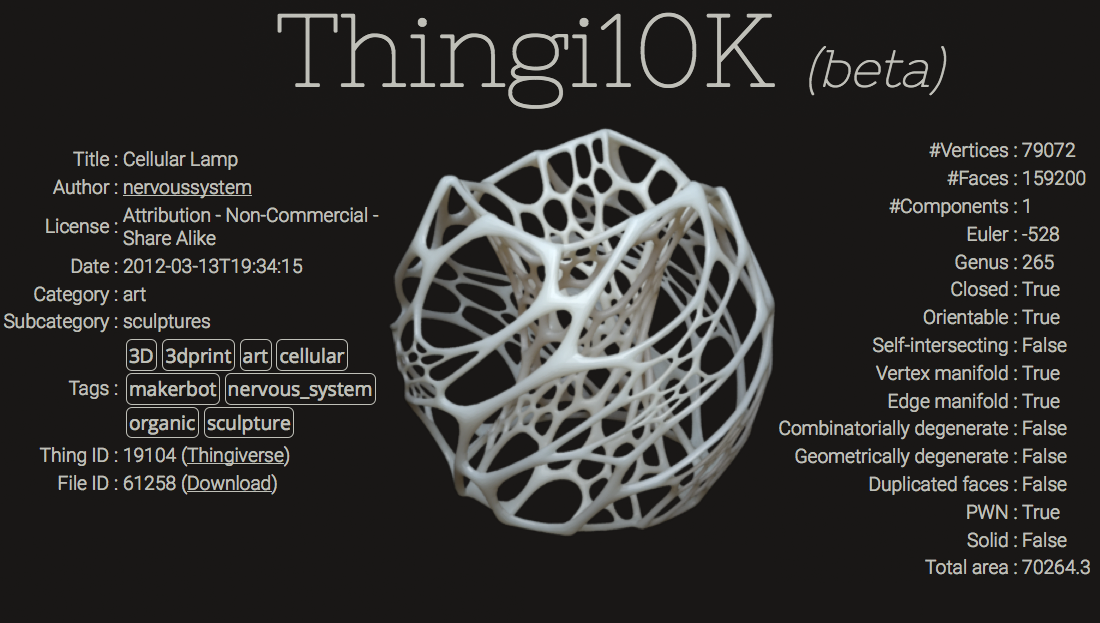}
\caption{Both contextual and geometric information
of each model are available on its model detail page.}
\label{fig:online_ui_detail}
\end{figure}

Our query interface is very useful in dissecting the dataset based on mesh
quality measures. For example, all single-component, manifold solid meshes
without self-intersection and degeneracies can be obtained with the query term
``\texttt{num component=1, is manifold, is solid, without self-intersection,
without degeneracy}''. All meshes satisfying these criteria are listed on the
result page (Figure \ref{fig:online_ui_result}).
We also provide an auto-generated python script to batch download results 
for custom search terms.

Users of our online query interface can view all contextual and geometry model
details (Figure \ref{fig:online_ui_detail}). In particular, we respect
the copyright of each model. On the model detail page, we clearly indicate the
original author and open source licence of each model. We also provide links to
the original Thingiverse pages where the raw data can be obtained.

\begin{figure*}[thbp]
\centering
\includegraphics[width=1.0\textwidth]{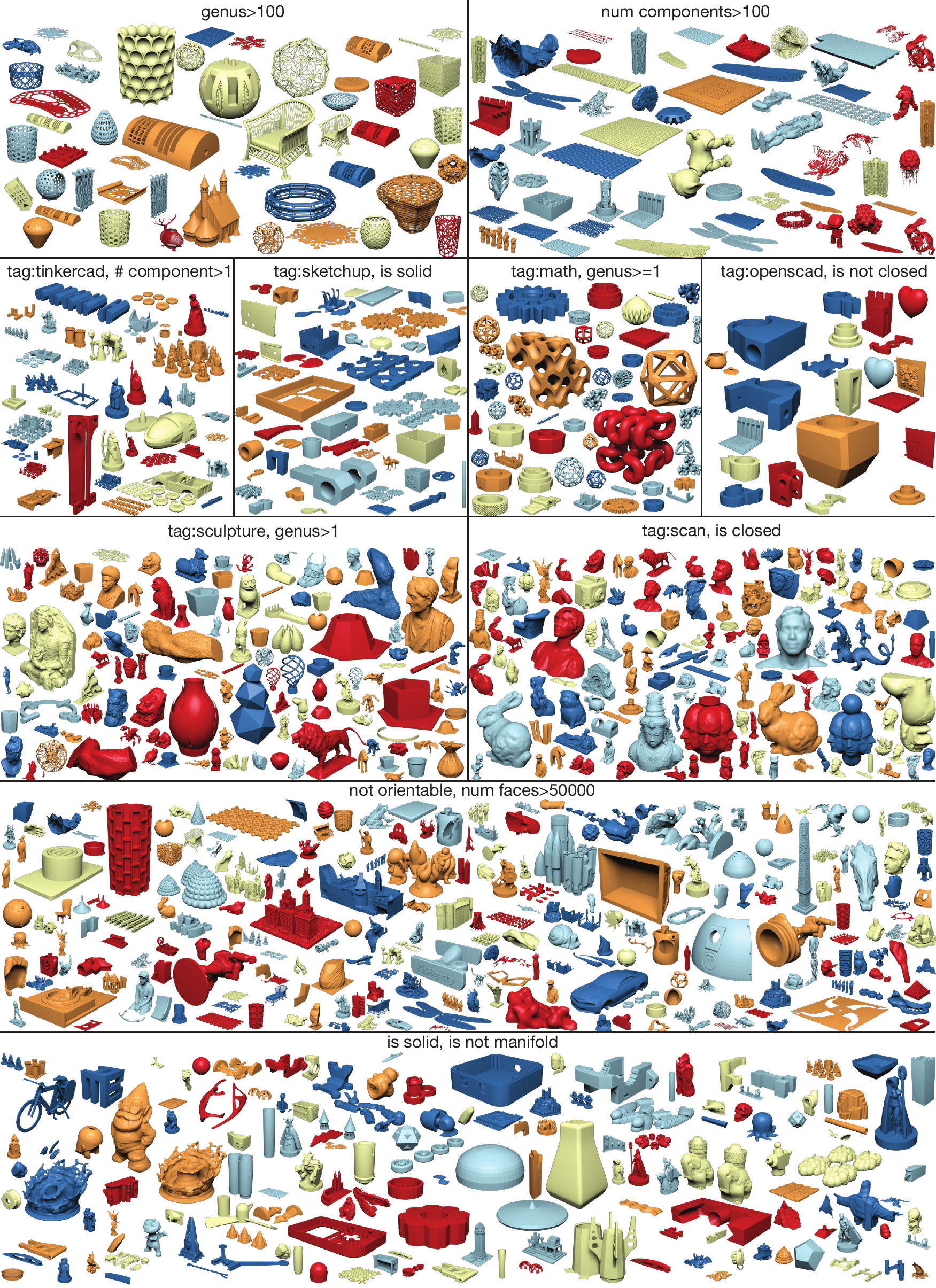}
\caption{
Our web interface returns subsets of the Thingi10K dataset via text queries.
}
\label{fig:query_examples}
\end{figure*}

To demonstrate the power of our online query interface, Figure
\ref{fig:query_examples} shows some interesting search results and the query used.

\section{Conclusion}
\label{sec:conclusion}

In this work, we present a large-scale annotated 3D dataset based on models used
in 3D printing applications.  Our dataset consists of 10,000 meshes crawled
systematically from Thingiverse.  We analyze both the contextual and geometric
information of our dataset and compare with two existing 3D model datasets.
Our analysis shows our data covers a large range of categories and provides a
balanced representation of real-world data in terms of mesh complexity and
quality.  The entire dataset and our analysis are freely available to the
public, and we provide a query interface to facilitate the exploration and
dissection of our dataset.

Our dataset could be used as input for stress-testing purposes as well as ground
truth for learning algorithms.  As for future work,
we plan to update and increase
the size of the dataset over time to reflect the fast-evolving nature of the 3D
printing community.
Specifically, we would like to include all
featured things from Thingiverse and add support for users to suggest additional
models for inclusion.
We hope our dataset and the accompanying analysis provide
an informative summary of 3D printing models and clarify the requirements for
geometry processing algorithms to be robust.

\section*{Acknowledgments}
%
We thank 
M.\ Campen, 
C.\ Tymms,
and J.\ Panetta
for early feedback and proofreading.
Funded in part by NSF grants  
CMMI-11-29917, IIS-14-09286, and IIS-17257.

\bibliographystyle{acmsiggraph}
\bibliography{references,10k_references}

\end{document}